# Magnetization dependent anisotropic topological properties in EuCuP

Jian Yuan,[1,†] Xianbiao Shi,[2,†] Hong Du,[3,†] Wei Xia,[1,6] Xia Wang,[1,7] Jinguang Cheng,[4] Baotian Wang,[2] Ruidan Zhong,[3,*] Shihao Zhang,[5,*] Yanfeng Guo[1,6,*]

[1] School of Physical Science and Technology, ShanghaiTech University, Shanghai 201210, China
[2] Institute of High Energy Physics, Chinese Academy of Sciences, Beijing 100049, China
[3] Tsung-Dao Lee Institute, Shanghai Jiao Tong University, Shanghai 201210, China
[4] Beijing National Laboratory for Condensed Matter Physics and Institute of Physics, Chinese Academy of Sciences, Beijing 100190, China
[5] School of Physics and Electronics, Hunan University, Changsha 410082, China
[6] ShanghaiTech Laboratory for Topological Physics, Shanghai 201210, China
[7] Analytical Instrumentation Center, School of Physical Science and Technology, ShanghaiTech University, Shanghai 201210, China

The correlation between magnetism and nontrivial topological band structure serves as a unique venue for discovering exotic topological properties. Combining magnetotransport measurements and first-principles calculations, we unveil herein that the hexagonal EuCuP holds topologically trivial state in the paramagnetic structure, while strong magnetization dependent anisotropic topological states in the spin-polarization structures. Specifically, it hosts a trivial topological state in the in-plane spin-polarization structure, while a Weyl semimetal state in the out-of-plane spin-polarization structure. Our scaling analysis suggests that the intrinsic Berry curvature in the spin-polarization structures can account for the observed large anisotropic anomalous Hall effect. First-principles calculations show that the magnetization and the spin-orbit coupling simultaneously play essential roles for the appearance of the four pairs of Weyl points in the out-of-plane spin-polarization structure. Our work therefore establishes in EuCuP the intimate relation between magnetism and the nontrivial topological states, which would be instructive for future study on this key issue of topological physics.

[†]The authors contributed equally to this work.
[*]Corresponding authors:

rzhong@sjtu.edu.cn

zhangshh@hnu.edu.cn,

guoyf@shanghaitech.edu.cn.

**INTRODUCTION**

The interplay between magnetism and nontrivial topological states has become one of the research frontiers in topological physics, because it can produce rich topological properties both in the bulk and surface states [1-21]. An explicit example is the Cr-doped topological insulator (TI) (Bi, Sb)$_2$Te$_3$ [6], in which the introduction of magnetism breaks the time reversal symmetry and opens a small Dirac mass gap in the topological surface state associated with the quantum anomalous Hall effect. To investigate the correlation between the two collective states, topological phases with strong spin-orbit coupling (SOC) and long-range magnetic order could serve as excellent platforms. In such systems, the applied external magnetic field can rotate the spins along different crystallographic directions and substantially influence the electronic band structure by the energy of even one order of magnitude larger than the traditional Zeeman splitting [23]. It therefore offers extraordinary opportunities to expose the tight relation between different magnetic structures and the corresponding topological electronic states. The van der Waals antiferromagnetic (AFM) topological insulators (TI) MnBi$_2$Te$_4$ [23], MnSb$_2$Te$_4$/(Sb$_2$Te$_3$)$_n$ [24, 25] and the layered EuCd$_2$X$_2$ (X = As, Sb) [26, 27] are such clear-cut examples which exhibit a topological phase transition from AFM TI to ferromagnetic (FM) Weyl semimetal (WSM) when the external magnetic field fully polarizes the spins along a specific crystallographic directions. More interestingly, the magneto-crystalline anisotropy and the relatively strong coupling between the local magnetization and the conduction electrons in cubic EuB$_6$ even can generate versatile magnetic topological phases along different crystallographic directions due to the varied magnetizations [28]. Regarding these, the exploration of topological phases with anisotropic magnetizations therefore allows for the achievement of in-depth insights into the correlation between magnetism and topological electronic states, as well as for the discovery of intriguing topological properties that could be used in next-generation spintronics.

The ternary Eu*TPn* compounds, where *T* denotes Cu, Ag, Au and *Pn* represents As, P, Sb, have captured vast attention in recent years owing to their rich magnetism and topological properties such as AFM Dirac semimetal, AFM nodal-line semimetal, etc [29-33]. Furthermore, it was predicted that the FM EuAgP hosts tunable topological states by rotating the spins along different crystallographic directions [34], which however still need experimental verification. Very recently, large anisotropic magnetotransport behaviors were observed in hexagonal EuCuP, while details of the correlation with the electronic band structure remain obscure [35]. It provides an opportunity to acquire a further understanding about the underlying physics.

In this work, we report on the study of EuCuP single crystals via comprehensive magnetotransport measurements and first-principles calculations. The results show that the strong correlation between the anisotropic magnetizations and the SOC effect results in anisotropic electronic topological band structures and hence largely anisotropic magnetotransport properties.

The EuCuP single crystals were grown by using tin as the flux. High purity europium rod (Alfa Aesar, 99.9%), copper superfine powder (Macklin, 99.9%), phosphorus (Aladdin, 99.999%) and tin pellet (Macklin, 99.99%) mixed in a molar ratio of 1:1:1:10 were placed into an alumina crucible and then sealed into a quartz tube in vacuum. The assembly was heated in a furnace up to 1100 °C within 15 h, kept at that temperature for 20 h, and then slowly cooled down to 700 °C at a temperature decreasing rate of 2 °C/h. The excess tin was immediately removed at this temperature by quickly placing the assembly into a high-speed centrifuge. The phase and quality of EuCuP single crystals were examined on a single crystal X-ray diffractometer equipped with a Mo K$\alpha$ radioactive source ($\lambda$ = 0.71073 Å). The detected diffraction patterns could be satisfactorily indexed on the basis of a BeZrSi-type structure (space group: $P6_3/mmc$, No. 194) with the lattice parameters $a = b$ = 4.122 Å, $c$ = 8.196 Å, and $\alpha = \beta$ = 90°, $\gamma$ = 120°. These values are very close to the ones reported previously [36].

Isothermal magnetizations at various temperatures between 2 and 35 K were measured on a commercial magnetic property measurement system from Quantum Design within the magnetic field range of 0 - 7 T. The temperature dependent magnetic susceptibility $\chi(T)$ was measured with the zero-field-cooling (ZFC) and field-cooling (FC) modes. Magnetotransport measurements, including the longitudinal resistivity, magnetoresistance, and Hall resistivity, were carried out in a commercial DynaCool Physical Properties Measurement System from Quantum Design. The longitudinal resistivity and magnetoresistance were measured in a four-probe configuration and the Hall effect was measured by using a standard six-probe method.

The first-principles calculations were carried out within the framework of the projector augmented wave (PAW) method [37, 38]. The generalized gradient approximation (GGA) [39] with the Perdew-Burke-Ernzerhof (PBE) [40] formula, as implemented in the Vienna *ab initio* simulation package (VASP) [41-43], was employed. For all calculations, the cutoff energy for the plane-wave basis was set as 550 eV, and the Brillouin zone (BZ) sampling was performed with a Γ-centered Monkhorst-Pack $k$-point mesh of 12 × 12 × 6 size, and the total energy difference criterion was set as $10^{-6}$ eV for self-consistent convergence. The GGA + $U$ ($U$ = 5 eV) scheme was utilized to consider the effect of Coulomb repulsion in the Eu 4$f$ orbital.

**RESULTS AND DISCUSSION**

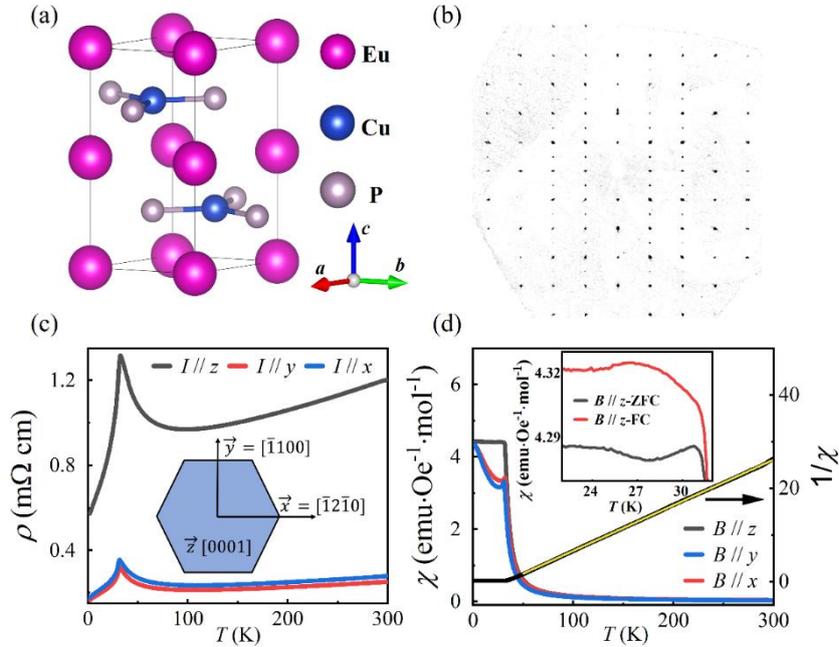

**Fig. 1.** (a) The crystal structure of EuCuP. (b) The XRD diffraction patterns in the reciprocal space

along (*h0l*). (c) The $\rho(T)$ with $I // x$, $I // y$, and $I // z$. Inset shows the crystallographic orientations. (d) Left: the FC $\chi(T)$ under $B = 0.05$ T. Right: reciprocal $1/\chi(T)$ with $B // z$. The inset presents an enlarged view of $\chi(T)$ around $T_C$.

The schematic crystal structure of EuCuP is shown in Fig. 1(a), which shows that the BeZrSi-type hexagonal structure is formed by alternately stacking Eu and Cu-P layers along the *c*-axis. The clean reciprocal diffraction pattern along (*h0l*) in Fig. 1(b) without other impurity spots indicates the high quality of our single crystals. The temperature dependent resistivity $\rho(T)$ with the electrical current *I* along *x*, *y* and *z* directions are presented in Fig. 1(c), where the layer-like stacking of the structure is along the *z* direction, as shown by the inset of Fig. 1(c). The $\rho(T)$ measured with $I // z$ is obviously larger than those with $I // x$ and $I // y$, indicating large electronic anisotropy. As temperature decreases, $\rho(T)$ peaks at around 31.5 K, which signifies the FM order. Fig. 1(d) presents the $\chi(T)$ measured at $B = 0.05$ T for $B // x$, $B // y$ and $B // z$, respectively. The Curie temperature is consistent with that observed in $\rho(T)$. As shown in Fig. 1(d), a fit to the reciprocal magnetic susceptibility $1/\chi(T)$ for $B // z$ by using the Curie-Weiss law $\chi(T) = C/(T - \theta_P)$ in the temperature range of 50 - 300 K, where *C* and $\theta_P$ denote the Weiss constant and Weiss temperature, respectively, gives the effective magnetic moment of ~ 8.95 $\mu_B$ and $\theta_P$ of ~ 32.0 K, indicating the FM magnetic ground state with a much larger effective magnetic moment than that expected for spin-only $Eu^{2+}$ ($Eu^{2+}$: $g\sqrt{S(S+1)}$ ~ 7.94 $\mu_B$ for $S = 7/2$). It could be interpreted by involving a substantial SOC effect in EuCuP and hence the contribution from the orbital moment. Besides, an anomalous behavior is observed in $\chi(T)$ with $B // z$, as shown by the inset of Fig. 1(d), where the FC curve shows a hump while the ZFC curve displays a concavity in the temperature range of 25 - 31.5 K. The behavior, which is invisible in $\chi(T)$ curves with $B // x$ and $B // y$, highly resembles the case of $EuB_6$ which exhibits a short-range magnetic order [44-47].

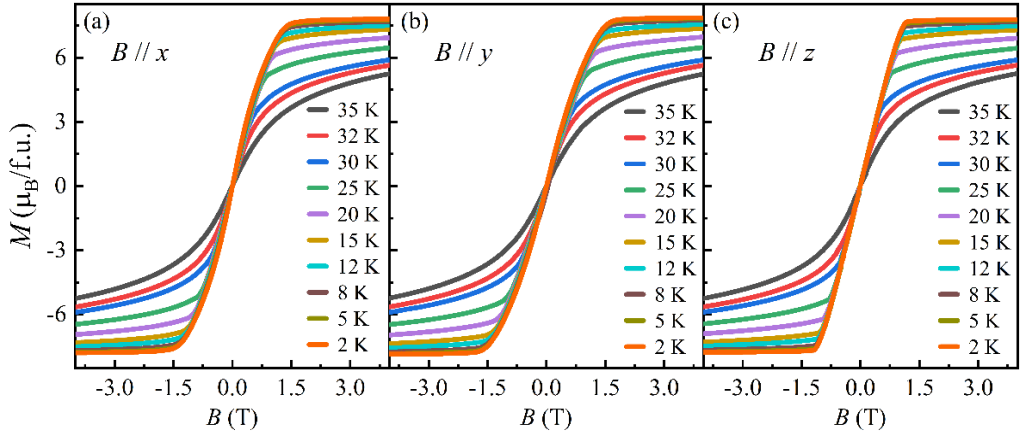

**Fig. 2.** (a) - (c) Isothermal magnetizations at various temperatures with $B \parallel x$, $B \parallel y$ and $B \parallel z$, respectively.

Fig. 2 presents the isothermal magnetizations $M(B)$ at different temperatures with the magnetic field $B$ along the $x$, $y$, and $z$ directions. The saturated magnetizations and the critical fields for $B \parallel x$, $B \parallel y$ and $B \parallel z$ are 7.82, 7.90, 7.81 $\mu_B$ and 1.44, 1.56, 1.15 T at 2 K, respectively, which indicates that the easy axis is the out-of-plane [0001] direction.

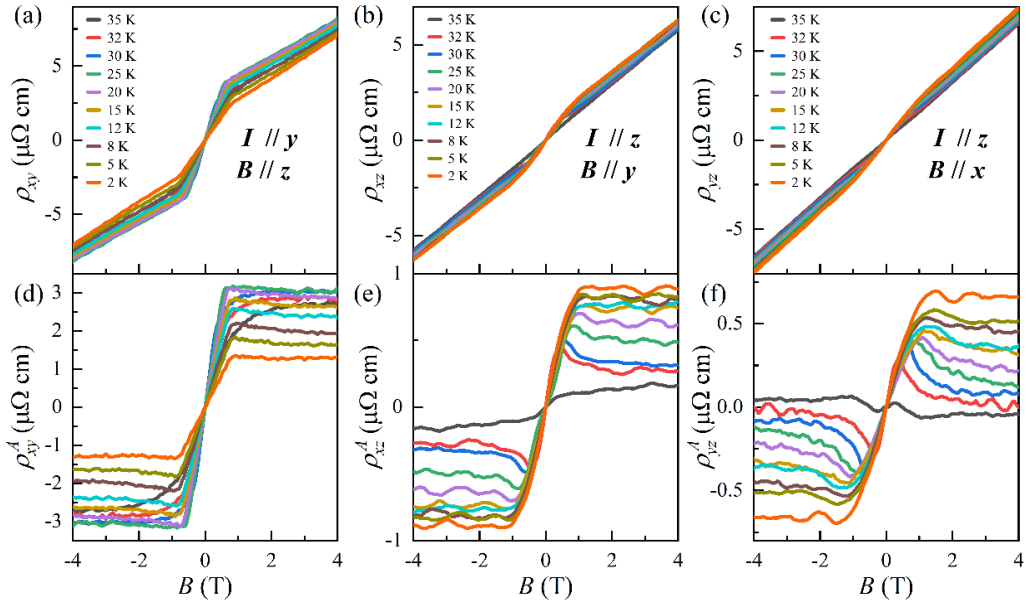

**Fig. 3.** The Hall resistivity versus magnetic field $B$ at different temperatures for (a) $I \parallel y$, $B \parallel z$, (b)

$I // z$, $B // y$ and (c) $I // z$, $B // x$, respectively. (d)-(f) are the corresponding anomalous Hall resistivity after subtracting the normal Hall resistivity.

The magnetotransport results with different $B$ directions are presented in Fig. 3. As $I$ and $B$ can be applied along $x$, $y$ and $z$ directions, the Hall resistivity is defined as $\rho_{ij} = \rho_{ij}(B_k)$ with $j // I$ and $k // B$. The nonlinear Hall resistivity $\rho_{xy}$ exposes the anomalous Hall effect (AHE) and the dominated hole-type carriers in EuCuP. The $\rho_{xy}$ can be expressed as $\rho_{xy} = \rho_{xy}^N + \rho_{xy}^A = R_0 B + 4\pi R_s M$, where $R_0$ is the normal Hall coefficient, and $R_S$ denotes the anomalous Hall coefficient. After subtracting the normal Hall resistivity, the anomalous Hall resistivity $\rho_{xy}^A$, $\rho_{xz}^A$ and $\rho_{yz}^A$ are obtained and presented in Figs. 3(d)-3(f). It is apparent that $\rho_{xy}^A$ is larger than $\rho_{xz}^A$ and $\rho_{yz}^A$. As temperature increases, $\rho_{xy}^A$ increases from 1.32 $\mu\Omega$ cm at 2 K to 3.14 $\mu\Omega$ cm at 25 K and then decreases to 2.70 $\mu\Omega$ cm at 35 K, while $\rho_{xz}^A$ and $\rho_{yz}^A$ have the largest values at 2 K and decrease with increasing temperature, which is reminiscent of the anomalous behavior seen in Fig. 1(d) and consistent with the result reported in the previous work [35]. However, our result shows clear difference in that when $B$ is along the in-plane direction, $\rho_{xz}^A$ and $\rho_{yz}^A$ are very small but not zero.

It is well known that topological semimetals with nontrivial Berry curvature host significant AHE. Taking the FM WSM $Co_3Sn_2S_2$ as an example, the intrinsic anomalous Hall conductivity (AHC) generated by the nontrivial Berry curvature even can reach 505 $\Omega^{-1}$ cm$^{-1}$ at 5 K [1, 2]. In $EuB_6$, the intrinsic AHC can reach 2050 $\Omega^{-1}$ cm$^{-1}$ when the magnetization $M // [111]$ at 2 K [28]. Since other extrinsic sources in a magnetic topological system, such as skew-scattering and side-jump contributions besides the nontrivial Berry curvature, can also produce AHE [48]. The analysis of the AHE should therefore be very careful. The so-called TYJ (Tian-Ye-Jin) model has been widely accepted to examine the intrinsic AHE [2, 28, 49-51]. Within this model, the anomalous Hall resistivity $\rho_{ij}^A$ can be expressed as $\rho_{ij}^A = a(M)\rho_{jj} + b(M)\rho_{jj}^2$, where $\rho_{jj}$ denotes the longitudinal resistivity, the first term comes from the extrinsic skew-scattering contribution, and the second term represents the intrinsic contribution from nontrivial Berry curvature and extrinsic side-jump contribution. It is easy to see that the relation of $\rho_{ij}^A/M\rho_{jj} = b\rho_{jj}$ can manifest the intrinsic AHE. As shown in Figs. 4(a)-4(c), $[\rho_{ij}^A/\rho_{jj}][M(0)/M(T)]$ for the three magnetic field directions are all linearly dependent on $\rho_{jj}$ in the spin-polarized states, suggesting the intrinsic AHE in EuCuP generated by the nonzero Berry curvature. The deviation from the linear curve in $\rho_{xy}^A$ when $\rho_{yy} >$ 0.185 mΩ cm is consistent with the anomaly seen in $\chi(T)$ with $B // z$. The above linear fitting can also give the coefficients $a$ and $b$. Generally, the AHC is defined as $\sigma_{ij}^A = -\rho_{ij}^A/\rho_{jj}^2$ (1) when $\rho_{ij}^A \ll \rho_{jj}$. The side-jump contribution can also be calculated by using the relation $\sigma_{ij,sj}^A = e^2/(ha)(\epsilon_{SOC}/E_F)$, in which $(\epsilon_{SOC}/E_F)$ is roughly 0.01 for ferromagnets, yielding the value of 9.37 Ω$^{-1}$ cm$^{-1}$ which is rather small. This component is therefore usually neglected in previous work [35]. However, we will show that it is nonnegligible when $B // xy$ in our case. By using the obtained coefficients $a$ and $b$, the intrinsic AHC can therefore be obtained.

As intrinsic AHC is linear to magnetization $M$ [52], we therefore can plot $\sigma_{In}^A \propto M$ as shown in Fig. 4(d), which shows a nice linear behavior. However, when further analyze the anisotropic intrinsic AHC, as summarized in Table I, for $\sigma_{ij,In}^A$ with $ij = xy$, $yx$, $xz$, $zx$, $yz$, and $zy$, the results show that $|\sigma_{xz,In}^A| \neq |\sigma_{zx,In}^A|$ and $|\sigma_{yz,In}^A| \neq |\sigma_{zy,In}^A|$ and the difference even reaches an order of magnitude, which apparently violates the Onsager's reciprocal relation. Because $\rho_{zz}$ is significantly larger than $\rho_{xx}$ and $\rho_{yy}$ as seen in Fig. 1(c), formula (1) can be modified as $\sigma_{ij}^A = -\rho_{ij}^A/\rho_{ii}\rho_{jj}$ (2) and the corresponding results are presented in Fig. 4(e), which still satisfy $\sigma_{In}^A \propto M$. The results suggest that the electronic structure anisotropy through analyzing the AHC must be carefully evaluated in such system, where the calculation of AHC should use formula (2) rather than (1), which was ever often overlooked in many references.

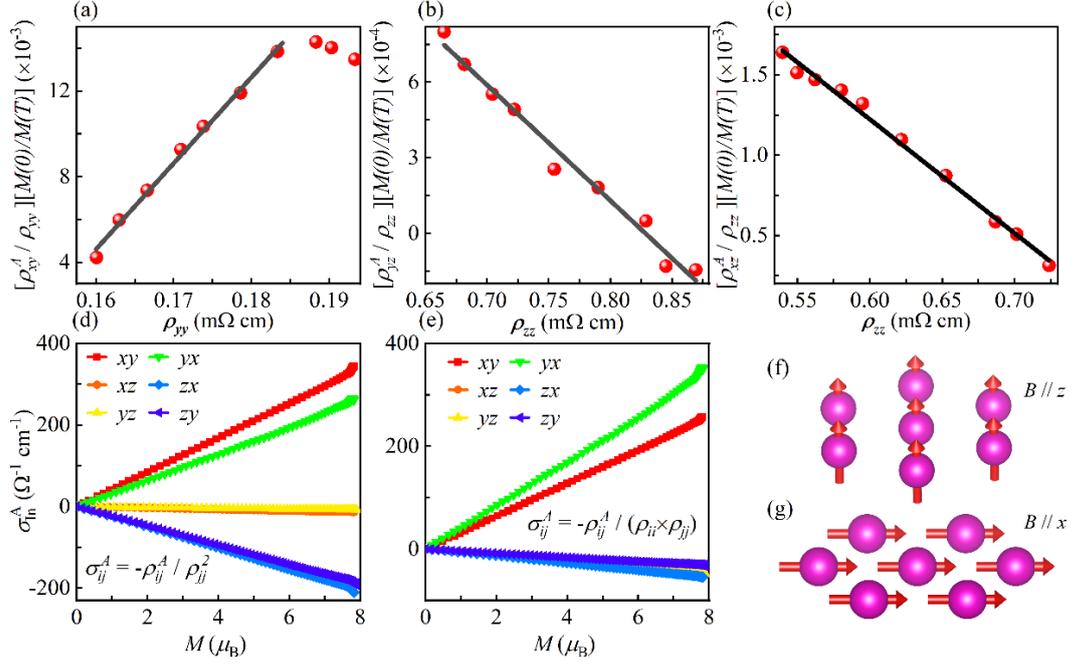

**Fig. 4.** (a) - (c) $[\rho_{ij}^A/\rho_{jj}][M(0)/M(T)]$ as a function of $\rho_{jj}$, where the solid line denotes the linear fit. (d) and (e) are intrinsic anomalous Hall conductivities with different directions at 2 K calculated by using the two formulas inserted in each figure, respectively. (f) and (g) are the schematic spin structures for $B // z$ and $B // x$, respectively.

The obtained intrinsic AHCs with $B // z$-axis are 347.17 and 256.09 $\Omega^{-1}$ cm$^{-1}$ for $I // x$ and $I // y$ at 2 K, respectively, which are much larger than $\sigma_{ij,sj}^A$, thus safely excluding the extrinsic contribution from the side-jump effect. The intrinsic AHC values are summarized in Table I, unveiling that magnetic field $B$ substantially affects the intrinsic AHC. When $B$ is along the out-of-plane direction, the intrinsic AHC is an order of magnitude larger than that for $B$ along the in-plane direction. However, the intrinsic AHCs for the in-plane and out-of-plane directions have opposite signs, which might be due to the existence of extrinsic effect. Since the coefficient $b$ is associated with intrinsic Berry phase and extrinsic side-jump, the side-jump could contribute an AHC of ~ 9.37 $\Omega^{-1}$ cm$^{-1}$, which is far smaller than the intrinsic AHC under a magnetic field along the $z$-axis but is close to the intrinsic AHC for a magnetic field within the $xy$-plane, indicating the side-jump effect cannot be neglected for the FM$x$ state. Figs. 4(f) and 4(g) are the schematic spin structures for $B // z$ and $B // x$, denoted as FM$z$ and FM$x$, respectively. A previous study on EuCuP suggested the presence and absence of Berry curvatures near Fermi level for $B // c$ and $B // ab$-plane, respectively, while without showing any details of the electronic band structure topology, thus leaving the correlation between magnetism and nontrivial topological states in EuCuP yet obscure [35]. Our magnetotransport results showing the magnetization-dependent anisotropic band structure topology in EuCuP expose the intimate relation between them.

**Table I**. The intrinsic AHCs with different directions calculated by using the two formulas (1) and (2).

| $\sigma^A_{ij,In}(\Omega^{-1}\text{cm}^{-1})$ | $xy$ | $yx$ | $xz$ | $zx$ | $yz$ | $zy$ |
|---|---|---|---|---|---|---|
| $\rho_{ii} = \rho_{jj}$ | 343.48 | 258.84 | -13.16 | -211.47 | -7.66 | -188.62 |
| $\rho_{ii} \neq \rho_{jj}$ | 256.09 | 347.17 | -44.25 | -55.06 | -49.05 | -30.15 |

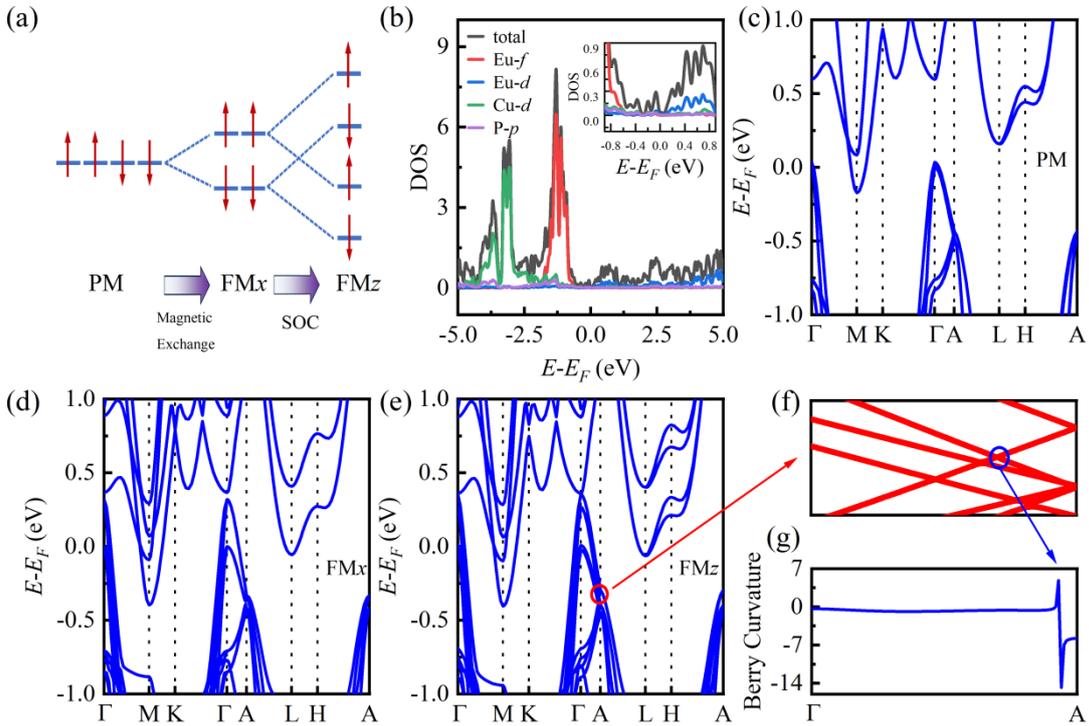

**Fig. 5.** (a) Schematic plot of different emergent band crossings with considering SOC and different magnetizations. (b) Density of states for EuCuP. (c)-(e) The band structures for (c) paramagnetic (PM), (d) FM$x$ and (e) FM$z$ states with SOC effect. (f) The zoom-in view for band structure of FM$z$

state and the blue circle emphasizes the band crossing. (g) The calculated Berry curvature along the Γ - A path.

To further investigate the origin for the large anisotropic intrinsic AHC, the electronic band structures of EuCuP were calculated, with the results shown in Fig. 5. Fig. 5(a) depicts the effects of magnetization and SOC effect on the band crossing in EuCuP. Fig. 5(b) presents the total and local density of states (DOS), where the low DOS near the Fermi level is consistent with the fact that EuCuP exhibits semimetallic conduction. The calculated band structures of the paramagnetic (PM) state are presented in Fig. 5(c), which shows both electron pockets and hole pockets around the Fermi level. The valence bands near the Fermi level are mainly contributed by the $d_{xy}$ and $d_{x^2-y^2}$ orbitals of Cu atoms and $p$ orbital of P atoms. Especially, the energy bands along high symmetric Γ - A path obey the four-fold degeneracy containing two-fold orbital degeneracy and two-fold spin degeneracy. When EuCuP system enters into the FM state, the two-fold spin degeneracy is lifted by the strong magnetic exchange as shown by the electronic structure of the FM$x$ state in Fig. 5(d). The two-fold orbital degeneracy is still kept in the FM$x$ state, because the SOC matrix elements $<m\sigma|L_+S_- + L_-S_+|n\sigma'>$ ($m,n = d_{xy}, d_{x^2-y^2}, \sigma, \sigma' =\uparrow, \downarrow$) are zero when the spins are along the in-plane $xy$ direction. However, when the spins are fully polarized along [0001], i.e. the FM$z$ state, the SOC matrix element becomes remarkable, because $<d_{xy}\sigma|L_ZS_Z|d_{x^2-y^2}\sigma> = i\lambda$ ($\sigma =\uparrow, \downarrow$), where $\lambda$ is the strength of SOC. Thus, in the FM$z$ state, the two-fold orbital degeneracy along high symmetric Γ - A path is split by strong SOC effect as shown in Figs. 5(e) and 5(f), which leads to band inversion as marked by the blue circle emerges along this Γ - A path, thus generating large Berry curvature near this position, as revealed by the calculated results shown in Fig. 5(g). Thus, a complete picture is achieved that the external magnetic field can induce a topological phase transition from a trivial state in FM$x$ structure to a Weyl semimetal in the FM$z$ structure through polarizing the spins along different directions, which is responsible for the large anisotropic intrinsic AHC. While the magnetization is along the easy axis [0001], i.e., the FM*z* state, four pairs of Weyl points that are slightly away from the high symmetric direction are identified, and their details are summarized in Table II. Based on the theoretical results which suggest a trivial topological state and the small AHC in FM*x* state, it is reasonable to attribute the AHC under magnetic field within the *xy*-plane to the extrinsic side-jump effect. Furthermore, the orientation of spins should also be carefully evaluated because it actually can affect the electronic band structure of a magnetic topological phase, as observed in Co doped FeSn [53 - 55]. However, the thickness of our measured crystals is around 0.2 mm, which suggests very strong bulk contribution and the effect of the surface modification should be very weak, which can only weakly influence the anomalous Hall effect in EuCuP. The theoretical results suggest that the correlation between magnetization directions and SOC effect plays a key role in determining the electronic band structure topologies in EuCuP.

**Table II**. The location and chirality of the calculated Weyl points in FM*z* state. Here Fermi level is set to zero for clarity.

| Weyl points (Ang$^{-1}$) | Chirality | Energy (eV) |
|---|---|---|
| ±(-0.03201, 0.11354, -0.01380) | ±1 | -0.035 |
| ±(-0.11281, -0.06513, 0.07436) | ±1 | -0.125 |
| ±(-0.17074, -0.09858, -0.31430) | ±1 | -0.487 |
| ±(0.00000, 0.20419, -0.31047) | ±1 | -0.494 |

**SUMMARY**

In summary, our study on single crystal EuCuP unveils anisotropic anomalous Hall effect with largely different anomalous Hall conductivities AHCs in the in-plane and out-of-plane directions. The AHCs with opposite signs indicate both the intrinsic and extrinsic contributions to the AHCs. It is also revealed that magnetization and SOC effect can induce a topological transition from a trivial semimetal in PM and FM*x* states to a Weyl semimetal in FM*z* state, which highlights the crucial role of magnetization in generating various nontrivial topological band structures and therefore provides useful insights into the correlation between magnetism and nontrivial topological states. The results would guide future design of new magnetic topological phases with tunable nontrivial topological states by magnetization. We would like to note that the SOC effect in Cu and P elements is somewhat weak. If Cu and P are substituted by using Ag and Sb or Bi, respectively, the stronger SOC would bring more substantial influences on the nontrivial topological band structure [56].

**ACKNOWLEDEMENTS**

The authors acknowledge the support by the National Natural Science Foundation of China (No. 92065201, 12304217, 12334008, 12025408 and 12374148), the National Key R&D Program of China (Grants No. 2023YFA1406100, 2021YFA1401600, and 2022YFA1402702). Y.F.G. acknowledges the open projects from State Key Laboratory of Functional Materials for Informatics (Grant No. SKL2022), CAS, and Beijing National Laboratory for Condensed Matter Physics. The authors also thank the support from the Analytical Instrumentation Center (SPST-AIC10112914) and the Double First-Class Initiative Fund of ShanghaiTech University.